\documentstyle[12pt]{article}  
\newcommand{\be}{\begin{equation}}
\newcommand{\ee}{\end{equation}}
\newcommand{\ba}{\begin{eqnarray}}
\newcommand{\ea}{\end{eqnarray}}
\newcommand{\no}{\nonumber\\}

\begin{document}
\begin{center}
 {\bf\Large{Hamilton - Jacobi treatment of
 front-form Schwinger model}}
\end{center}
\begin{center} {\bf Dumitru Baleanu}\footnote[1]{ On leave of absence from 
Institute of Space Sciences, P.O BOX, MG-23, R 76900
Magurele-Bucharest, Romania,
E-mail: dumitru@cankaya.edu.tr}
and
{\bf Yurdahan G\"uler}\footnote[2]{E-Mail:~~yurdahan@cankaya.edu.tr}
\end{center}
\begin{center}
Department of Mathematics and Computer Sciences, Faculty of Arts
and Sciences, Cankaya University-06530, Ankara , Turkey
\end{center}
\begin{abstract}
    The Hamilton-Jacobi formalism was applied to quantize 
the front-form Schwinger model. The importance of the surface term is 
discussed in detail. The BRST-anti-BRST symmetry was analyzed within
Hamilton-Jacobi formalism.  
\end{abstract}

\newpage

 \section {Introduction}

  Quantum electrodynamics in one-space and one-time dimension with 
massless charged fermion is known as the Schwinger model.
  It is the one of the very few models of field theory which can be solved 
analytically \cite{schgeneral1, schwinger, schgeneral2}.
 The Fock-space content of the physical states depends crucially on the 
coordinate system and on the gauge. It is only in the front form that a 
simple constituent picture emerges \cite{berko} as well as it is an 
important example of the type of simplification that  we hope will occur 
for QCD in physical space-time. The investigations of similar model with 
massive fermion and for non-abelian theory,  where the fermion is in the 
fundamental and adjoint representation, revealed that many properties are 
unique to the Schwinger model \cite{schwingeroriginal}.
 During last years light-cone quantization of quantum field theory has 
emerged as a promising method for solving problems in the strong coupling 
regime \cite{schwingerreview}.
 The front-form Hamiltonian and BRST formulation  \cite{usa}  and recently 
the extended Hamiltonian formalism of the pure  space-like axial gauge 
Schwinger model \cite{schwingerjaponia} were investigated. 
 The Schwinger model in the front form possesses a set of three
first-class constraints \cite{dirac1949}. In the instant form it has two
first-class constraints \cite{schwinger0}.        

  Hamilton-Jacobi formalism was subjected to various
investigations during the last years. 
 The formalism was applied
for strings and p-Branes  \cite{hjstringsbrane} and for strongly
coupled gravitational systems \cite{hjsalopek} and in addition  the
quantum  Hamilton-Jacobi formulation was obtained from the equivalence
principle \cite{hjquantum}. A new method of quantization of the system with constraints based on the 
$Carath{\acute e}odory$'s equivalent Lagrangians method \cite{car} was 
initiated \cite{g5,g6}  by one of us.
   
 Recently this formalism was generalized to the singular systems
with higher order Lagrangians and to systems which have  elements
of the Berezin algebra \cite{p11, p12, p13}. The quantization of
the systems with constraints was investigated using this approach
\cite{gb1,gb2,gb3}.
 Recently, the connection between the  integrability  conditions and
Dirac's consistency conditions \cite{dirac} were established \cite{p13}. 
 Importance of the surface terms \cite{mark} was analyzed as well as the 
relation  between Batalin-Fradkin-Tyutin \cite{fad} and Hamilton-Jacobi  
formalism \cite{gb2}.     
 Even more recently the formalism was applied to
investigate  the nonholonomic-constrained systems with second-class
constraints \cite{hong} as well as Proca's model \cite{hong1}.    

 The  advantage  of the Hamilton-Jacobi formalism
is that we have no difference between first and second class
constraints and we do not need gauge-fixing term because
the gauge variables are separated in the processes of constructing  an
integrable system of total differential equations.
 In addition the action provided by the formalism can be
used in the process of path integral quantization method of the 
constrained systems.

 The main aim of this paper is to investigate  the quantization of the
front-form Schwinger model and its BRST extension  using Hamilton-Jacobi 
formalism. 
\\
\\

The plan of the paper is the following:

In sec. 2 the Hamilton-Jacobi formalism is presented.
In sect. 3 the front-form Schwinger model and its BRST extension are 
analyzed using Hamilton-Jacobi
formalism. In sec. 4 conclusions are given. 

\section{Hamilton-Jacobi formalism}
Let us assume that the Lagrangian L is singular and the Hessian 
supermatrix has rank n-r.
The Hamiltonians to start with are

\be\label{doi} H_{\alpha}^{'}=H_{\alpha}(t_{\beta},q_{a},p_{a})
 +p_{\alpha}, \ee where $\alpha,\beta=n-r +1,\cdots,n$,$a=1,\cdots
n-r$. The usual Hamiltonian $H_0$ is defined as

\be\label{unu} H_{0}=-L(t,q_{i},{\dot q_{\nu}},{\dot q_{a}=w_{a}})
+p_{a}w_{a} + {\dot
q_{\mu}}p_{\mu}\mid_{p_{\nu}=-H_{\nu}},\nu=0,n-r+1,\cdots,n. \ee
which is independent of $\dot q_{\mu}$.Here $\dot
q_{a}={dq_{a}\over d\tau}$,where $\tau$ is a parameter.
 The equations of motion are obtained as total differential equations
in many variables as follows

\ba\label{(pq)}
&dq_{a}&=(-1)^{P_a +P_a P_\alpha} {\partial_{r} H_{\alpha}^{'}\over\partial
p_{a}}dt_{\alpha},
dp_{a}=-(-1)^{P_{a}P_{\alpha}}{\partial_{r} H_{\alpha}^{'}\over\partial
q_{a}}dt_{\alpha},\cr
&dp_{\mu}&=-(-1)^{P_{\mu}P_{\alpha}}{\partial_{r} H_{\alpha}^{'}\over\partial
t_{\mu}}dt_{\alpha}, \mu=1,\cdots, r , \ea

\be\label{(z)} dz=(-H_{\alpha} + (-1)^{P_a + P_a 
P_\alpha}p_{a}{\partial_{r}
H_{\alpha}^{'}\over\partial p_{a}})dt_{\alpha}, \ee where
$z=S(t_{\alpha},q_{a})$ and $P_{i}$ represents the parity  of $a_{i}$.

Determination of the degrees of freedom in the Hamilton-Jacobi formulation 
really needs elaboration.
Although one starts with a system of n degrees of freedom the theory 
forces to reduce it due to the integrability conditions. In other words, 
variations of constraints may cause new constraints and again
their variations leads one to a degree of freedom which is less than n. If 
the variations vanish identically, of course degrees of freedom do not 
change.

\section{The front-form Schwinger model}
The Lagrangian density of the Schwinger model in one-space, one -time dimension
is
\be\label{lagra}
L^{'}=\psi\gamma^{\mu}(i\partial_{\mu}+gA_{\mu})\psi -{1\over 
4}F_{\mu\nu}F^{\mu\nu}
\ee
where $F^{\mu\nu}={\partial^{\mu}A^{\nu}-{\partial^{\nu}A^{\mu}}}$.

The bosonised version has the following Lagrangean density 
\cite{schwinger0, refsch}.

\be\label{boso}
L^{''}= -{1\over 4}F_{\mu\nu}F^{\mu\nu}  +{1\over 2}(\partial_{\mu}\phi)^2
-g \epsilon^{\mu\nu}\partial_{\mu}\phi  A_{\nu} ,
\ee
where $g^{\mu\nu}= diag (1, -1)$, $\epsilon^{01}=-\epsilon^{10}=1$,
or in the component form
(\ref{boso}) becomes

\be\label{treil}
L^{'''}={1\over 
2}({\dot\phi}^2-{\phi^{'}}^2)+g({\phi^{'}A_{0}}-{\dot\phi}A_{1})
+{1\over 2}({\dot A_{1}}-{\dot A_{0}^{'}})^2.
 \ee

The light-cone coordinates are defined as
\be
x^{\pm}={1\over\sqrt{2}}(x^{0}\pm x^{1}).
\ee

The Lagrangian (\ref{treil}) in the light-cone coordinate form becomes
\be\label{la3}
L=(\partial_{+}\phi)({\partial_{-}}\phi)+g(\partial_{+}\phi)A^{+}-
g(\partial_{-}\phi)A^{-}+
{1\over 2}({\partial_{+}A^{+}}-{\partial_{-}}A^{-})^2.
\ee
Here $A_{\pm}={1\over\sqrt{2}}(A_{0}\pm A_{1})$ and 
$\partial_{\pm}\phi={1\over\sqrt{2}}({\dot\phi}\pm\phi^{'})$.

>From (\ref{la3}) we calculate the canonical momenta as
\be\label{canonici}
\Pi^{+}=0,
\Pi^{-}={\partial_{+}A^{+}-{\partial_{-}}A^{-}},
\Pi={\partial_{-}\phi}+gA^{+},
\ee
where $\Pi^{+},\Pi^{-}$ and $\Pi$ are conjugate to $A^{-}, A^{+}$ and
$\phi$ respectively.
>From (\ref{canonici}) the Hamiltonian densities to start with are
\be\label{hdoi}
H_{1}^{'}=\Pi^{+},
H_{2}^{'}=\Pi-{\partial_{-}\phi-gA^{+}}
\ee
The canonical Hamiltonian density is
\be\label{canonical}
H_{c}={1\over 2}(\Pi^{-})^2 +\Pi^{-}(\partial_{-}A^{-})+g(\partial_{-}\phi)A^{-}.
\ee
In Hamiltonian-Jacobi formalism we have , at this stage, three Hamiltonians
densities

\ba\label{haaa}
&H_{1}^{'}&=\Pi^{+},
H_{2}^{'}=\Pi-{\partial_{-}\phi-gA^{+}},\cr
&H_{0}^{'}&=p_0+{1\over 2}(\Pi^{-})^2 +\Pi^{-}(\partial_{-}A^{-})+
g(\partial_{-}\phi)A^{-}
\ea
The gauge variables are $A^{-}$ and $\Phi$ and the independent one is
$A^{+}$.
The equations of motion corresponding to (\ref{haaa}) are
\ba
&dA^{+}&={\partial H_{0}^{'}\over\partial\Pi^{-}}d\tau=(\Pi^{-}
+{\partial_{-}A^{-}})d\tau,\cr
&d\Pi^{-}&=-{\partial H_{0}^{'}\over\partial A^{+}}d\tau- {\partial 
H_{2}^{'}\over\partial A^{+}}d\Phi =
gd\Phi.
\ea
>From $dH_{1}^{'}=0$ we find  that
\be\label{hami3}
H_{3}^{'}={\partial_{-}\Pi^{-}-g{\partial_{-}\Phi}}=0.
\ee
Constraint (\ref{hami3}) is not suitable for Hamilton-Jacobi but if we 
consider it as a density  the contribution to the action is zero.
At this point we can see the main difference between Dirac's formulation
and Hamilton-Jacobi formalism for fields.
To keep the physical interpretation in Hamilton-Jacobi formulation
we must have all the constraints in the form of $p_{\alpha}+ H_{\alpha}$.
In our concrete case the constraints are not in the form required but they
are first class.
The system is integrable on the surface of constraints, in other words when
$H_{1}^{'}=H_{2}^{'}=H_{3}^{'}=0$.
For this model the reduced phase-space is suitable in the Hamilton-Jacobi
formalism.  In this case we have
\be
H_{0}^{'}=p_{0}+{1\over 2}(\Pi^{-})^2.
\ee

\subsection{BRST Invariance}

The main aim of this section is to find a way to relate BRST 
transformation \cite{brst}
to the Hamilton-Jacobi formulation.
The first problem is the Lagrangian to start with.
The starting point is the BRST invariant Lagrangian  \cite{usa}

\ba\label{labrst}
&L^{'}&=\Pi^{-}{\partial_{+}A^{+}}+ \Pi_{u}{\partial_{+}u}+
\Pi_{v}{\partial_{+}v}-{1\over 2}(\Pi^{-})^2 -\Pi^{-}(\partial_{-}A^{-})
-gA^{-}(\partial_{-}\phi)\cr
&+&(\partial_{-}\phi+gA^{+})\partial_{+}\phi.
\ea
obtained from the total Hamiltonian 

\be
H_{T}={1\over 2}(\Pi^{-})^2 +\Pi^{-}(\partial_{-}A^{-})+
g(\partial_{-}\phi)A^{-} +\Pi^{+}u +(\Pi-\partial_{-}\Phi-gA^{+})v.  
\ee

Here $\Pi_{u}$ and $\Pi_{v}$ are the momenta corresponding to Lagrange's 
multipliers u and v.

The BRST and anti- BRST symmetries for Schwinger model have the forms
\ba\label{be}
&\delta\phi=0&,\delta A^{+}={\partial_{-}c},\delta A^{-}=\partial_{+}c,
\delta u={\partial_{+}\partial_{+}c},\delta v=0,\cr
&\delta\Pi=g\delta_{-}c&,\delta\Pi^{+}=0,\delta\Pi^{-}=0,\delta\Pi_{u}=0,
\delta\Pi_{v}=0\cr
&\delta c=0&, {\delta}{\bar c}=0,\delta b=0,
\ea

\ba\label{antibe}
&{\bar\delta}\phi=0&,{\bar\delta} A^{+}=-{\partial_{-}{\bar 
c}},{\bar\delta} A^{-}=-\partial_{+}{\bar c},
 {\bar\delta} u=-{\partial_{+}\partial_{+}{\bar c}},{\bar\delta} v=0,\cr
&{\bar\delta}\Pi=-g{\bar\delta}_{-}{\bar 
c}&,{\bar\delta}\Pi^{+}=0,{\bar\delta}\Pi^{-}=0,{\bar\delta}\Pi_{u}=0,
{\bar\delta}\Pi_{v}=0\cr
&{\bar\delta}{\bar c}=0&, {\bar\delta}{\bar c}=0,{\bar\delta} b=0,
\ea
           
where $c$, ${\bar c}$ are Grassmann variables, b 
is a bosonic variable and in addition  $\delta^2=0$, 
${\bar\delta}^2=0$. 

The supercharges corresponding to (\ref{be}) and (\ref{antibe}) are

\be
Q=\int dx^{-}[ic(\partial_{-}\Pi^{-}-g\partial_{-}\phi)-
i\partial_{+}c(\Pi^{+}
+\Pi-{\partial_{-}\phi}-gA^{+})],
\ee

\be
{\bar Q}=\int dx^{-}[-i{\bar c}(\partial_{-}\Pi^{-}-g\partial_{-}\phi)+
i\partial_{+}{\bar c}(\Pi^{+}
+\Pi-{\partial_{-}\phi}-gA^{+})],
\ee  

After adding a  gauge-fixing term to  (\ref{labrst}) we obtain \cite{usa}
                                                      
\be\label{lagbrst}
 L_{BRST}=L^{'}+\delta({\bar c}({\partial_{+}A^{+}}+{1\over 2}b-gA^{+}
 +\Pi))
 \ee
or 

\ba\label{laabrst} 
&L_{BRST}&=\Pi^{-}{\partial_{+}A^{+}}+ \Pi_{u}{\partial_{+}u}+
\Pi_{v}{\partial_{+}v}-{1\over 2}(\Pi^{-})^2 -\Pi^{-}(\partial_{-}A^{-})\cr
&-&gA^{-}(\partial_{-}\phi) +(\partial_{-}\phi+gA^{+})\partial_{+}\phi
+{1\over 2}b^2 +b(\partial_{+}A^{-}-gA^{+}+\Pi)\cr
&+&(\partial_{+}{\bar c})(\partial_{+}c).
\ea
Since the Lagrangian (\ref{laabrst}) is degenerate on the extended 
phase-space, we can apply Hamilton-Jacobi formulation.

 The Hamiltonian densities corresponding to (\ref{laabrst}) are
\ba\label{has}
&H_{1}^{''}&=\Pi^{+}-b,
H_{2}^{''}=\Pi_{c}-{\partial_{+}{\bar c}},
H_{3}^{''}=\Pi_{{\bar c}}-{\partial_{+} c},
H_{4}^{''}=\Pi-{\partial_{-}\Phi}-gA^{+},\cr
&H_{0}^{''}&=p_0+{1\over 2}(\Pi^{-})^2 
+\Pi^{-}(\partial_{-}A^{-})+g(\partial_{-}\phi)A^{-}
+\Pi_{c}\Pi_{{\bar c}}\cr
&-& \Pi^{+}(\Pi-gA^{+})-
{1\over 2}(\Pi^{+})^{2}
\ea

We mention that in (\ref{has}) all Hamiltonians are BRST- anti- BRST 
invariant if $\delta p_{0}= {\bar\delta p_{0}}=0$.  

The corresponding equations of motion for independent variables are

\be
d\Pi^{-}=-g\Pi^{+}dx^{+}-gd\Phi,
dA^{+}=(\Pi^{-}+{\partial_{-}A^{-}})dx^{+}
\ee

The variations of $H_{1}^{''},H_{2}^{''},H_{3}^{''},H_{4}^{''}$ gives

\be\label{beu}
d\Pi^{+}=db,
d\Pi={\partial _{-}d\Phi}+gdA^{+}
\ee

\be\label{ghosts} 
d\Pi_{c}={\partial _{+}d{\bar c}},
d\Pi_{{\bar c}}={\partial _{+}d{c}},
\ee

>From (\ref{ghosts}) we obtain 
\be
{\partial_{+}(\partial_{+}c)=0},
{\partial_{+}(\partial_{+}{\bar c})=0},    
\ee

On the other hand $d\Phi=-g\Pi^{+}dx^{+}-gd\Phi$, then 
$\Pi^{+}=0$. From (\ref{beu}) we obtain  $b=0$.
The equation of motion corresponding to $\Pi^{+}$ is 
\be
d\Pi^{+}={\partial_{-}\Pi^{-}-g{\partial_{-}\Phi}}
\ee
As a consequence  a new constraint appears and it is considered as a 
new Hamiltonian density

\be
H_{5}^{''}=\partial_{-}\Pi^{-}-g{\partial_{-}\Phi}.
\ee

Solving  $H_{2}^{''}=0,H_{3}^{''}=0 $  we obtain  c 
and ${\bar c}$ as
\be\label{gosts}
c(x^{+})=Ax^{+}+B, {\bar c}(x^{+})=A^{'}x^{+}+B^{'}.
\ee

Let $\mid\psi>$ be the physical state of the model.
All Hamiltonians from (\ref{has}) annihilate the physical state.
We mention that this condition is the same as
\be
Q\mid\psi>=0,\\
{\bar Q}\mid\psi>=0.
\ee 

Taking into account (\ref{ghosts}) we obtain that the physical 
states are described by

\ba
&\Pi^{+}\mid\psi>=0&,
(\Pi-{\partial_{-}\phi}-gA^{+})\mid\psi>=0,\cr
&(\partial_{-}\Pi^{-}-g{\partial_{-}\phi})\mid\psi>&=0.
\ea 
The result is the same as obtained in \cite{usa}.
 
\section{Conclusions}

Despite of many attempts to clarify the Hamilton-Jacobi formalism 
for the systems with constraints a lot of problems remained unsolved.
In this paper we analyzed the front-form Schwinger model  with 
Hamilton-Jacobi and the results are the same as those obtained with 
Dirac's formalism.
Using the consistency conditions we found three first class constraints 
but one of them is not in a form 
required by Hamilton-Jacobi formalism. As a consequence in this case 
we cannot calculate the action. To solve the problem we find the 
reduced phase-space.
We mention that $H_{3}^{'}$ becomes a total divergence in 
Hamilton-Jacobi formalism.
This model is an example of a constrained system such that all the 
constraints are first class but its physical interpretation from 
Hamilton-Jacobi point of view is missing.
In the second part of the paper we started with a BRST invariant 
Lagrangian. Since the Lagrangian is singular on the extended phase-space 
we apply Hamilton-Jacobi formalism.
All  Hamiltonian densities are BRST- anti- BRST invariant 
provided that $\delta p_{0}= {\bar\delta p_{0}}=0$. The equations for 
c and ${\bar c}$ were obtained and the results are in agreement with 
those obtained in literature. 

\section {Acknowledgments}
The authors would like to thank M. Henneaux and S.T. Hong for 
reading the manuscript and for encouragement.
This work is partially supported
by the Scientific and Technical Research Council of Turkey.

\section{Appendix}

Let us consider the Lagrangian $L(q, {\dot q})$ be an even function  of
N variables $q^{I}$ that are elements of Berezin algebra.
The variation of $S=\int L dt$  leads us to the equations of motion 

\be
{\delta_r S\over\delta q^i} ={\partial_r L\over \partial q^i}- {{d\over dt}
{\partial_r L\over \partial \dot{q^i}}}=0,
\ee

The Hamiltonian becomes 
\be
H= p_i{\dot q^i}- L, 
\ee
where the momenta are defined using the right derivatives 
\be
p_i={\partial_r L\over\partial{\dot q^i}},
\ee

and Hamilton's equations are 
\be
{\dot q^i}={\partial_l H\over\partial p_i}=(-1)^{P_{(i)}}{\partial_r H\over\partial p_i},
{\dot p^i}={-\partial_r H\over\partial q_i}=-(-1)^{P_{(i)}}{\partial_l H\over\partial q_i}.
\ee

The Berezin bracket is defined as
\be
\{F,G\}_B={\partial_r F\over\partial q^i}{\partial_lG\over\partial p_i}
\no
-(-1)^{P_{(F)}P_{(G)}}{\partial_r G\over\partial q^i}{\partial_l F\over\partial p_i}
\ee
and has the following properties
\be
\{F,G\}_{B}=-(-1)^{P_{(F)}P_{(G)}}\{G,F\}_{B},
\ee
\be
\{F,GK\}_{B}=\{F,G\}_B K+ (-1)^{P_{F}P_{G}}\{F,K\}_{B},
\ee

\ba\label{jacobi}
(-1)^{P_{(F)}P_{(K)}}\{F,\{G,K\}_{B}\}_{B}
&+& (-1)^{{P_{(G)}P_{(F)}}}\{G,\{K,F\}_{B}\}_{B}\cr
&+& (-1)^{P_{(K)}P_{(G)}}\{K,\{F,G\}_{B}\}_B = 0.
\ea
Here (\ref{jacobi}) represents the Jacobi's identity.

\end{document}